\documentstyle[12pt, epsbox]{article}

\begin{document}

{\bf\Large
Search for an evidence of Fermi acceleration for SNR in a time dependence of metal abundance}

Satoko Osone
{\it Matsudo 1083-6 A-202, Matsudo, Chiba 271-0092, Japan}\\

\section{Abstract}
 {\small  
A time dependence of a metal abundance of supernova remnant (SNR) have been found by Hughes et al. (1998). 
We interpret this time dependence is due to Fermi acceleration.
If this interpretation is true, there is a metallicity dependence of this time scale. However, we could not find  an evidence of Fermi acceleration.
} 

\section{Introduction}

Fermi acceleration has been accepted as an acceleration mechanism of particles to a high energy.
The energy spectra of cosmic rays is represented as a power law of an energy (Simpson 1983), which is expected from Fermi acceleration.
The origin of cosmic ray has been considered as SNRs. 
Koyama et al. (1995) confirmed an existence of a shell for SN1006 with {\it ASCA}. The limb brighten is dominated by a thermal process in the shock-heated interstellar material and/or supernova eject.
And, they detected non-thermal emission from a limb and suggested an existence of  electrons of energy above 200 TeV.
Tanimori et al. (1998) detected TeV gamma emission from SNR SN 1006.
This emission is interpreted as comptoned cosmic micro background of 2.7K by high energy electrons.
These results suggest that Fermi acceleration is occurred in SNRs.

Hughes et al. (1998) found there is an anti correlation between an age and a metal abundance with seven SNRs in LMC.
If an ejector of star is observed as emission lines, a metal abundance derived from emission lines depends on both a mass of an original star and a type of an explosion. If interstellar medium(ISM) is observed, a metal abundance is constant in same LMC.
Hughes et al. (1998) suggested that metal abundance show ISM for seven SNRs.
They interpreted that this relation indicates the presence of newly produced SN eject in at least some of the remnants of the sample.

Here, we interpreted that this relation suggest the decrease of low energy metallicity that is shown as emission line by Fermi acceleration.
This  time scale of decrease will depend on the metallicity according to Fermi acceleration.
Therefore, we search for metallicity dependence of this time scale as a evidence of Fermi acceleration.

\section{Data and Result}

We use fitting results of Hughes et al. 1998.
There are two kinds of a metal abundance and an age which are derived from two models, full electron and ion temperature equilibration ( model 1 ) and column equilibration ( model 2 ). 
Here, we assume that acceleration conditions as a magnetic fields are same for all SNRs in LMC.

The rate of an acceleration is expressed as (Gaisser 1990)
\begin{equation}
\frac{dE}{dt} = \frac{\xi E}{T_{cycle}}
\end{equation}
Here, $\xi$ is an energy gain ratio per one encounter.
The cycle time of an acceleration $T_{cycle}$ is written as(Gaisser 1990)
\begin{equation}
T_{cycle}=\frac{4}{c}(\frac{D_1}{u_1} + \frac{D_2}{u_2})
\end{equation}
Here, $u_1,u_2$ are velocities of a flow in an up stream and a down stream respectively  and
$D_1,D_2$ are coefficients of a diffusion in an up stream and a down stream respectively.
A coefficient of a diffusion can not become smaller than a Larmor radius.
Hence, 

\begin{equation}
D_{min} = \frac{r_L c}{3}\sim\frac{1}{3}\frac{Ec}{ZeB}
\end{equation}
Here, $Z$ is a charge of particles and $B$ is a magnetic field.
Therefore, by applying $D_1=D_2=D_{\rm min}$, we obtain 
\begin{eqnarray}
\frac{dE}{dt} &\sim& Z\alpha
\end{eqnarray}
Here, $\alpha$ is an independent constant of an energy.
This relation means that particle of high metallicity is accelerated more rapidly than  that of low metallicity.

We fitted a relation between an age $t$ and a metal abundance $N$ of seven SNRs with $N=a\times t^{b}$ for each metallicity. Here, $a,b$ are free parameters. We show fitting results in table 1. The error is 1 sigma statistical error. 
We show the relation between an age and a metal abundance of seven SNRs for a metal S of model 1 in figure 1.
Most fitting results show a decrease of metal abundance with an age.
This is consistent with Hughes et al. (1998).

We make a relation between the fitted index $b$ and charge $Z$.
If there is a Fermi acceleration, we may obtain anti correlation between $b$ and $Z$.
We fitted this relation with a constant and obtain a reasonable statistical result.
Therefore, we could not find an evidence of Fermi acceleration.

\vspace{10pt}
We thank prof.Makishima for discussion.

\begin{table*}
\caption{The fitted results of a relation between a metal abundance $N$ and an age $t$ for each metallicity with $N=a\times t^{b}$. $a, b$ are free parameters.}
\begin{center}

\vspace{10pt}
\begin{tabular*}{18cm}{ccccccc}   \hline\hline
metallicity & O & Ne & Mg & Si & S & Fe \\ \hline
\multicolumn{7}{c}{model 1}\\ \hline
b  & $-$0.22$^{-0.20}_{+0.20}$ &  $-$0.29$^{-0.16}_{+0.16}$ & 0.28$^{-0.25}_{+0.25}$ & $-$0.65$^{-0.35}_{+0.35}$ & $-$0.46$^{-0.29}_{+0.29}$ & $-$0.40$^{-0.23}_{+0.23}$ \\
$\chi^2$/d.o.f  & 4.70/5 & 4.87/5 & 6.50/5 & 6.82/3 & 3.50/3 & 7.48/5 \\ \hline
\multicolumn{7}{c}{model 2}\\ \hline
b &  $-$0.55$^{-0.42}_{+0.42}$ & $-$0.37$^{-0.28}_{+0.28}$ & 0.21$^{-0.36}_{+0.36}$& $-$0.21$^{-0.26}_{+0.26}$ & $-$0.46$^{-0.40}_{+0.40}$ & $-$0.60$^{-0.22}_{+0.22}$ \\
$\chi^2$/d.o.f  & 6.10/5 & 2.58/5 & 3.58/5 & 4.99/5 & 2.96/3 & 6.94/5 \\ \hline
\end{tabular*}
\end{center}
\end{table*}

\begin{table}
\caption{The fitted results of a relation between the index $b$ and a metallicity $Z$ for each model.}
\begin{center}
\vspace{10pt}
\begin{tabular*}{8.5cm}{ccc}   \hline\hline
 model & constant & $\chi^2/d.o.f$ \\ \hline
model 1 &  $-$0.25$\pm$0.09 & 7.04/5 \\
model 2 & $-$0.36$\pm$0.12 & 4.30/5 \\
\end{tabular*}
\end{center}
\end{table}

\begin{figure}[htb]
 \begin{center}
\psbox[height=8.5cm,width=8.5cm]{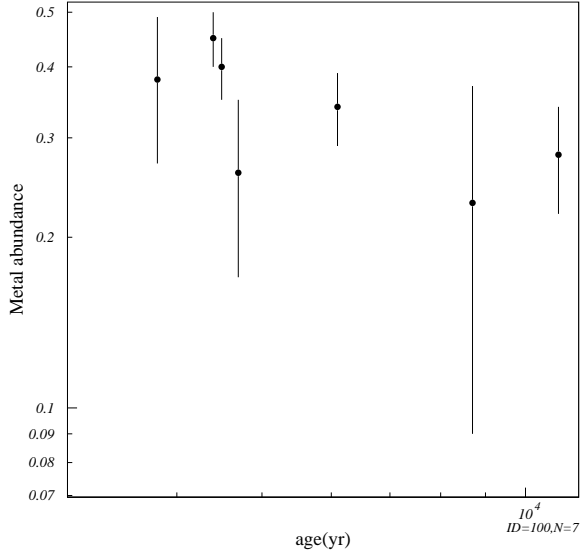}
 \end{center}
\caption{The relation between an age and a metal abundance of seven SNRs for a metallicity S of model 1}
\end{figure}

\begin{figure}[htb]
 \begin{center}
\psbox[height=8.5cm,width=8.5cm]{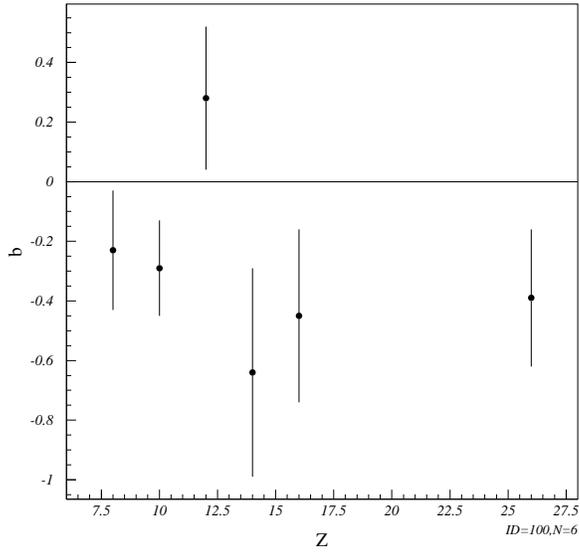}
\psbox[height=8.5cm,width=8.5cm]{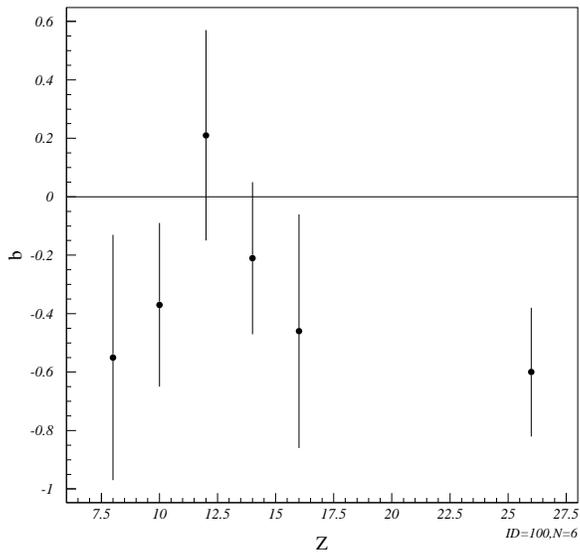}
 \end{center}
\caption{The relation between $Z$ and the index $b$ for model 1(left) and model 2(right)}
\end{figure}

\vspace{\baselineskip}

\section{References}

\begin{itemize}
\setlength{\itemsep}{-1.5mm}
\setlength{\itemindent}{-8mm}
\item[] Gaisser, T.K., 1990, {\it Cosmic Rays and Particle Physics}, Cambridge University press. 
\item[]  Hughes,J.P., Hayashi, I.\& Koyama.K., 1998, ApJ, 505, 732
\item[] Koyama, K., Petre, R., Gotthelf, E.V., Hwang, U., Matsuura, M., Ozaki\& M., Holt, S.S., 1995, Nature, 378, 255
\item[]  Simpson, J.A., 1983, Ann.Rev.Nucl.Part.Sci., 33,323
\item[] Tanimori, T, Hayami, Y.,Kamei, S., Dazeley, S.A., Edwards, P.G., Gunji, S., Hara, S., Hara, T., et al., 1998, ApJ, 497, L25 
\end{itemize}

\end{document}